# Multifunctional Liquid Metal Lattice Materials with Recoverable and Reconfigurable Behaviors

Fanghang Deng, Quang-Kha Nguyen, Pu Zhang[1]

*Department of Mechanical Engineering, State University of New York at Binghamton, Binghamton, NY 13902, US*

**Abstract:** Multifunctional lattice materials are of great interest to modern engineering including aerospace, robotics, wave control, and sensing. In this work, we proposed and developed a new class of functional lattice materials called liquid metal lattice materials, which are composed of liquid metals and elastomer coatings organized in a co-axial way. This hybrid design enables the liquid metal lattice materials with a thermal-activated shape memory effect and hence a variety of intriguing functionalities including recoverable energy absorption, tunable shape and rigidity, deployable and reconfigurable behaviors. The fabrication of liquid metal lattice materials is realised by a hybrid manufacturing approach combining 3D printing, vacuum casting, and coating. The mechanical properties and functionalities of these liquid metal lattice materials are studied via experimental testing. We expect that this new class of lattice materials will greatly expand the current realm of lattice materials and lead to novel multifunctional applications.

---

[1] Corresponding author. Email: pzhang@binghamton.edu

# 1 Introduction

Lattice materials [1] are composed of repetitive unit cells with artificially designed geometry to achieve weight reduction and/or desirable functionalities. In recent few years, lattice materials have been categorized into a more general class of materials, namely, architected materials [2]. In addition, since lattice materials have considerable overlap with mechanical metamaterials [1,3], these two terms are sometimes used interchangeably by researchers. Early development of lattice materials focused on simple structures like honeycomb structure, mesh, and foam [4,5] due to the limitation of conventional manufacturing technologies. In contrast, the last decade has witnessed an eruption of lattice materials with complicated geometry [6,7], hierarchical structure [8,9], gradient design [10,11], multimaterial [12,13], and multifunctionality [1,12,14], owing to the rapid development and adoption of the additive manufacturing (or 3D/4D printing) technology [15]. Right now, this area is still growing rapidly and we envision that more and more novel lattice materials will be developed in the near future. In addition, promising applications of lattice materials will also be realized in aerospace, robotics, biomedical, sensing, wave control, among others.

Multifunctional lattice materials exhibit functionalities beyond conventional load-bearing usage and are usually enabled by multiphysical fields such as thermal, electrical, magnetic, and acoustic ones. Among the most intriguing functionalities are, for instance, tunable shape and rigidity, shape memory, reconfigurability, sensing and actuation, wave filtering, color change, and so on. Most of the multifunctional lattice materials are fabricated by additive manufacturing. We classify multifunctional lattice materials into four categories. (1) *Thermal*. Many literature work falls into this category such as lattice materials with shape memory effect [14,16,17], tailored thermal expansion [12], and designed heat conductivity [18]. In particular, by using shape memory polymers [14,16,17], researchers are able to fabricate lattice materials with tunable properties, recoverable shapes, deployable and reconfigurable behaviors. (2) *Electrical*. Recently, Zheng and his coworkers [19] manufactured piezoelectric lattice materials with potential applications in sensing and actuation. On the other hand, Hopkins and coworkers [20] designed and fabricated a few electrically-activated lattice metamaterials by using hybrid manufacturing methods. (3) *Magnetic*. Most of magnetic lattice materials are based on elastomers filled with magnetic particles. By applying an external magnetic field, the lattice materials can be either deformed or reconfigured [21,22]. Researchers also fabricated magneto-rheological lattice materials [23] with tunable rigidity by using 3D printing and liquid filling. (4) *Acoustic*. Lattice materials can be used to manipulate acoustic fields. The additive manufacturing technology has been used frequently to fabricate such kind of acoustic lattice metamaterials in recent years [21]. We classify the lattice metamaterials controlling elastic waves [24] in this category as well. Very recently, multifunctional lattice materials activated by more than one physical fields are also emerging, e.g. by combining the thermal field with acoustic/magnetic field.

Overall, most published works on multifunctional lattice materials have focused on tunable shape and rigidity, programmability, and reconfigurability to date, possibly driven by the demanding applications in soft robotics [25], soft implants, tunable wave control, and deployable components, among others. The materials being employed are mostly shape memory polymers [14,16,17] due to their intrinsic flexibility and the ease of fabrication. However, it is known that shape memory polymers exhibit some limitations [26–28], e.g. low stiffness, slow response speed, low thermal conductivity, wide transition temperature, lack of precisely tunable stiffness and transition temperature, etc. In contrary, shape memory alloys [29] such as NiTi does not suffer from these limitations but is lack of flexibility for many modern applications. Therefore, there is a demanding need to develop shape memory materials bridging the gap between traditional shape memory polymers and shape memory alloys for multifunctional lattice materials usage.

In this work, we introduce the design and manufacturing of a new class of shape memory lattice materials, namely, liquid metal lattice materials. These lattice materials are composed of a liquid metal core and an elastomer coating layer so that the whole lattice material exhibits an extrinsic shape memory effect originating from the phase transition of liquid metals, which will be explained in Section 2. Since direct additive manufacturing of such lattice materials is still challenging, we adopt a hybrid manufacturing approach by combining 3D printing, vacuum casting, and coating together, which is introduced in Section 3. The developed liquid metal lattice materials exhibit a variety of intriguing functionalities including recoverable energy absorption, shape and rigidity tuning, deployable and reconfigurable behaviors, which will be introduced in Section 4. Finally, Section 5 will be dedicated to further discussions of the liquid metal lattice materials.

## 2    Design and Mechanism

The underlying mechanism of shape memory effects is based on phase transition [26,27]. Liquid metals, if used as transition phases, will offer a promising solution to achieve shape memory effects that overcome many limitations of shape memory polymers. The term liquid metal is used to classify pure metals and alloys with melting points, $T_m$, not far from room temperature [30]. Some commonly used liquid metals include Galinstan ($T_m = $ -19°C), EGaIn ($T_m = 16$°C), Ga ($T_m = 30$°C), and Field's metal ($T_m = 62$°C). These liquid metals experience a solid-liquid phase transition by heating the solid to a melting point $T_m$, and some of them are solids at room temperature. As a result, liquid metals like Ga and Field's metal can bring about shape memory effects if they are embedded in an elastomer matrix or skeleton.

Here we propose a new class of shape memory lattice materials by coating a liquid metal lattice material with an elastomer shell, as shown in Fig. 1. Without this elastomer shell skeleton, the liquid metal phase will collapse once melted, and hence the elastomer coating is crucial to maintain the structural integrity of liquid metal lattice materials. The shape memory mechanism of this liquid metal lattice material is illustrated in Fig. 1 by taking a single unit cell of a Kelvin lattice as an example. In Fig. 1, the permanent shape of this unit cell (Fig. 1a) is dictated by its elastomer shell skeleton; and the temporarily fixed shape (Fig. 1b) can be programmed by melting, deforming, and then freezing the embedded liquid metal lattice structure. In this work, we use Field's metal to fabricate the lattice materials. Upon heating above $T_m$, the permanent shape of the liquid metal lattice material is restored since the silicone shell skeleton will recover to its original shape, as shown in Fig. 1c. Precisely, the liquid metal lattice materials display an extrinsic shape memory effect, as opposed to the intrinsic effect of shape memory polymers and alloys [26,29]. Note that the coating layer must exhibit an elastic behavior so the permanent shape can be restored. If any inelastic deformation in the coating layer exists, the shape memory response will be impaired.

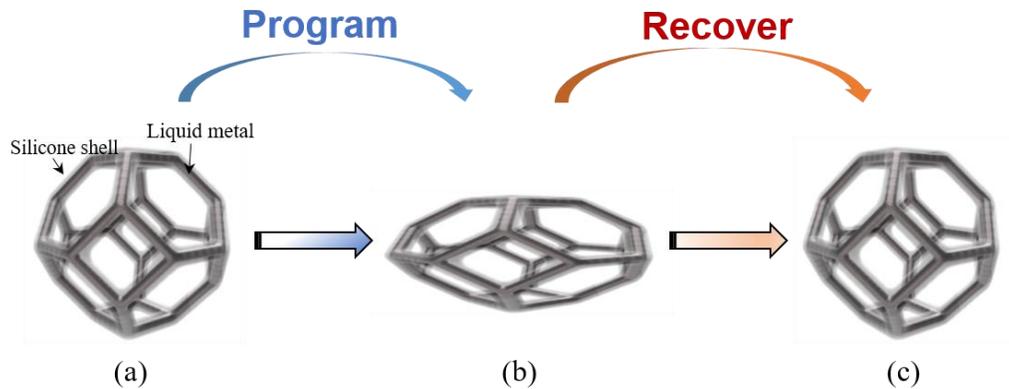

**Figure 1**. Shape memory mechanism of liquid metal lattice materials. (a) A Kelvin lattice unit cell in its original (permanent) shape. The unit cell is composed of a liquid metal core and a silicone coating. (b) The liquid metal lattice is heated, deformed, and then cooled down to a programmed state, i.e. a temporarily fixed shape. (c) The permanent shape of the unit cell can be recovered by melting the liquid metal core to release the stress.

The current literature on the shape memory effect of liquid metal-based materials is limited. At present, much of the work on liquid metals is based on EGaIn and Galinstan [31–36], which are liquids at room temperature and not ideally suited for shape memory applications. There has not been a lot of work reported on the shape memory effect induced by the phase transition of liquid metals, in particular Ga and Field's metal. Shan and Majidi [37] fabricated a composite structure with tunable rigidity by using liquid metals and elastomer in 2013. Later on, Shepherd and his coworkers [38] fabricated a Field's metal sponge and tested the stiffness change upon heating and the reconfigurable behavior. In 2018, Bartlett and Thuo [39] synthesized liquid metal composites by mixing PDMS with micro-particles of Field's metal. Bartlett and Thuo observed similar reconfigurable behavior as in Shepherd's work [38]. In 2019, Dickey and coworkers [40] reported an ultrastretchable shape memory fiber with a rubber coating and Ga core. They discovered that the response speed of this liquid metal-based shape memory fiber (2-6 s) is much faster than conventional shape memory polymers. Very recently, Kramer-Bottiglio and coworkers [41] synthesized silicone- and epoxy-based liquid metal composites with shape memory effect. Almost meanwhile, Hopkins and his coworkers [42] published an article on the shape memory effect of metamaterials fabricated by packing together arrays of Ga-filled elastomer spheres. These early studies demonstrate that the solid-liquid phase transition of liquid metal inclusions introduces shape memory behavior in the liquid metal-polymer composites and structures. However, there is no report of works on liquid metal lattice materials so far.

## 3   Hybrid Manufacturing Process

We propose a hybrid manufacturing process to fabricate the liquid metal lattice materials since direct additive manufacturing of these materials remains challenging in the current stage. The fabrication process mainly consists three steps: 3D printing, vacuum casting, and coating.

Figure 2 shows a detailed procedure of the proposed hybrid manufacturing process. Each of the three major steps is introduced below in details.

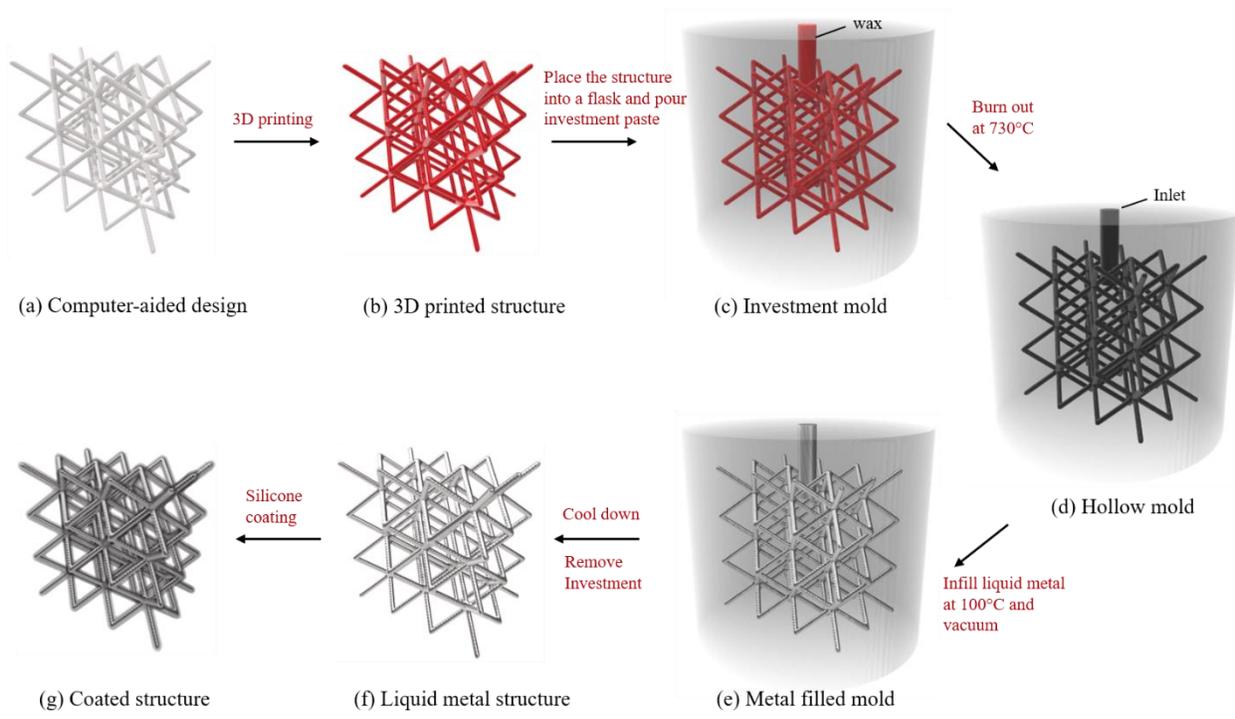

**Figure 2**. Hybrid manufacturing process of liquid metal lattice materials. The fabrication process consists of three steps: (a-b) 3D printing; (c-e) Vacuum casting; and (f-g) Coating.

**STEP 1: 3D printing.** The 3D printing is done by using a commercial DLP (digital light projector) printer. A lattice structure is built in a computer-aided design software as shown in Fig. 2a. The output STL file is sliced with a layer height of 50 μm and sent to a DLP printer (Hunter, Flashforge Inc., China). A castable blend resin (Fun To Do Inc., Netherlands) is used to print the lattice structure with default processing parameters recommended by the Flashforge Hunter 3D printer. The 3D printing process is usually successful without any visible defects or failures. Note that the resin needs to be stirred and filtered frequently to guarantee the printing quality. After 3D printing, the lattice structure is detached from the building plate, pruned to remove support structures, and rinsed thoroughly by isopropyl alcohol (99.5%, EMD Millipore Co.) and then deionized (DI) water. Finally, the rinsed lattice structure is dried in air (Fig. 2b).

STEP 2: Vacuum casting. (1) The 3D printed lattice structure is then fixed on to a sprue base by using casting wax and placed into a casting flask (Fig. 2c). The investment paste is prepared by mixing investment powder (Plasticast with Bandust, Ransom & Randolph Inc.) and DI water with a weight ratio of 100:52. This brand of investment powder consists of silica, Calcium sulfate, and some amount of quartz. It is water soluble after burning so ideally suited for the lattice structure casting process. After the investment paste is prepared well, it is poured into the casting flask slowly and degassed in a vacuum chamber for 2-3 min until boiling to remove trapped air bubbles (Fig. 2c). The investment mold is rested at room temperature for 2-5 h to harden and dry before burning. (2) The burning process follows a flash mode recommended by the castable resin provider. A furnace is preheated to 600℃ at beginning. The casting mold is placed into the furnace and heated to 730℃ at a rate of 2.5℃/min and then dwelled for 2 h. After that, the furnace is cooled down to 280℃ at a rate of 2.2℃/min and held at this temperature for 4-6 h before the investment molds are moved out. This burning process will burn out the sacrificial polymer lattice resulting in a hollow mold (Fig. 2d) for the following casting process. (3) The hollow mold is further cooled in air to 100℃ until the casting process is started. A piece of Field's metal ($T_m = 62℃$, Rotometals Inc.) is melted and poured into the hollow mold. Afterwards, the metal-filled mold (Fig. 2e) is placed into a vacuum chamber to assist the filling process. This vacuum process will significantly improve the casting quality of the lattice structures and reduce undesirable defects. It is recommended that this vacuum process should last for at least 5 min. (4) After the casting process, the metal-filled mold is cooled down to room temperature and cleaned by water jet to remove the investment powder. Finally, a liquid metal lattice structure can be obtained (Fig. 2f).

STEP 3: Coating. The fabricated liquid metal lattice structure (Fig. 2f) is coated with a thin layer of elastomer coating as the shell skeleton (Fig. 2g). We have screened over ten types of coating resins. The one chosen in this work is the silicone conformal coating (DOWSIL 1-2577 Low VOC, Dow Inc.). This silicone coating is cured by humidity in the air and has a short tack-free time (~ 6 min). The silicone resin is diluted by 5 % of toluene (Honeywell Inc.) to reduce the viscosity. The dip coating method is used to fabricate the lattice materials. Note that spray coating won't work since the coating is usually much thicker on the edge beams than interior beams. For the dip coating process, the lattice structure is immersed into the diluted silicone resin, drawn out slowly, and cured in air for 10 min. After each coating cycle is completed, we proceed to the next cycle, and a total of 15 layers are coated on the liquid metal lattice structure with a total thickness about 350 μm. Finally, the coated liquid metal lattice structures are cured for 12 h at room temperature and then 5 h at 50℃ in a furnace.

From the explanation above, it is conceived that the fabrication of liquid metal lattice materials has taken the advantage of the additive and conventional manufacturing methods. With the 3D printing technology, sacrificial molds of complicated geometry can be readily fabricated for investment casting usage. We will show a variety of liquid metal structures fabricated by this hybrid method in Section 4 below.

Even though this hybrid manufacturing process is demonstrated to be very successful, it still encounters with some technical challenges and limitations below.

(1) At first, typical casting defects are observed including cold shut along some beams and very thin fins between beams. The casting defects are strongly affected by the porosity of the investment mold, which is controlled by tailoring the water ratio. Empirically, reducing the porosity of the mold will increase its strength and eliminate the fins that are caused by fracture of the mold. However, low porosity will usually induce cold shut defects since the vacuum process is not effective enough to remove air thoroughly. Thus, the casting process needs to be optimized to reduce the defects.

(2) Manufacturing accuracy is also an issue though not critical. Overall, we have found satisfactory accuracy in all of our fabricated liquid metal structures. The manufacturing accuracy is mainly affected by the 3D printing and casting processes. The DLP printer normally produces very accurate part but sometimes distortion can be observed at locations with delicate geometry or insufficient support structures. The casting accuracy is affected by the shrinkage of the investment mold. For example, the as-fabricated liquid metal beams are usually slightly thicker than the designed sizes.

(3) Defects in the coating layers may also be found if inappropriate coating materials are chosen. We have tried a few different coating resins. Typical coating defects are invisible pin holes that will cause leakage of the liquid metal after melting. Moreover, the strength and durability of the coating materials are also crucial to the mechanical performance and functionalities of the liquid metal lattice materials. We have found that conformal coatings that are originally designed for electronic packaging are among the best candidates for the current coating process.

## 4   Multifunctional Behaviors

The designed and manufactured liquid metal lattice materials exhibit an extrinsic shape memory effect. This effect will enable the lattice materials several intriguing functionalities like recoverable energy absorption, tunable shape and rigidity, deployable and reconfigurable behaviors, among others. We will introduce these functionalities below for some of liquid metal lattice materials we fabricated and tested. All the CAD models are displayed in Section S1 of the SI.

## 4.1 Recoverable energy absorption

Recoverable energy absorption materials are appealing since they avoid catastrophic failure and promote sustainability. In the literature, lattice materials with recoverable energy absorption ability are usually based on shape memory polymers [14], bi-stable unit cell design [43], and granular-based metamaterials [44]. These lattice materials usually have a relatively low energy absorption capacity since their stiffness is low. In contrast, liquid metals like Field's metal exhibit much higher stiffness than polymers and hence will dissipate a lot more energy. On the other hand, compared to shape memory alloys, liquid metal lattice materials have much larger reversible strain range due to their unique hard-soft integrated design. Therefore, liquid metal lattice materials have high potential to be used as recoverable protection or cushion layers in engineering.

To prove this concept, we fabricate three types of liquid metal lattice materials: Kelvin, BCC (body centered cubic), and honeycomb structures, as shown in Fig. 3a-c. The dimensions of the three lattice structures are identical, i.e. $26.3 \times 26.3 \times 18.2$ mm$^3$. The beams are designed as 0.9 mm thick, whereas the as-fabricated diameter is within 1-1.1 mm, slightly thicker than the designed size. The actual volume ratios of Field's metal are 8.69%, 5.90%, and 8.33% for these three lattice structures, respectively.

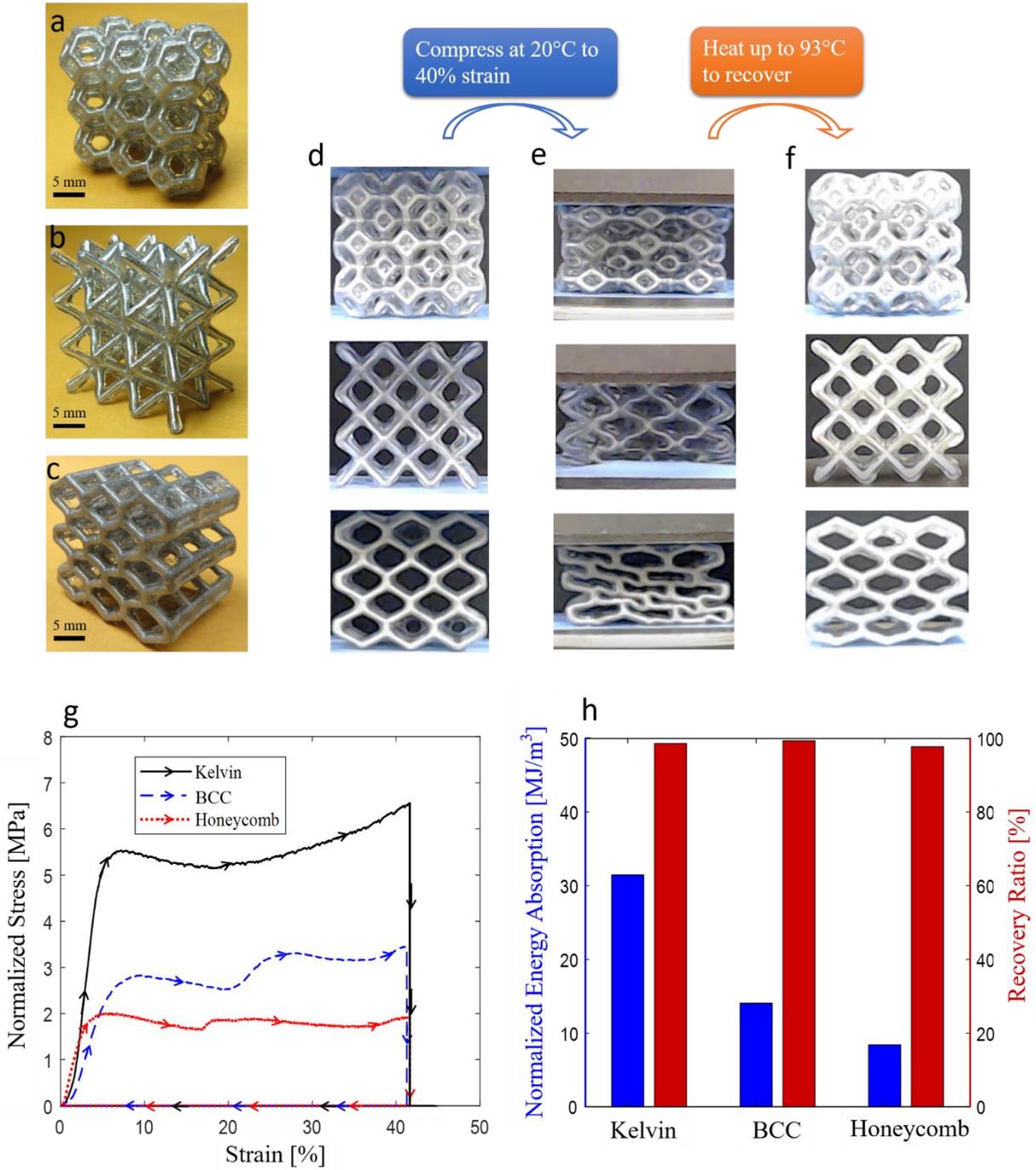

**Figure 3**. Recoverable energy absorption behavior of liquid metal lattice materials. (a) Kelvin lattice. (b) BCC lattice. (c) Honeycomb structure. (d) Original shapes of lattice materials before compression. (e) Lattice materials after compression at room temperature. The compression process experiences large plastic deformation and energy dissipation. (f) Lattice materials restore their original shapes after melting the liquid metal core. (g) Comparison of the normalized stress-strain curves of the three lattice structures. (h) Comparison of the normalized energy absorption and recovery ratios of the three lattice structures.

The energy absorption capacity and recoverability of these three types of liquid metal lattice materials are tested. As shown in Fig. 3d-e, the lattice materials are compressed to 40% engineering strain at a loading speed of 0.1 mm/s on a load frame (MTS 858 System, 5 kN load cell). The compression testing is conducted at room temperature. After the compression testing, the load is removed and lattice materials remain to be in their deformed shapes due to the plastic deformation of the metal. A typical stress-strain curve of the Field's metal can be found in Section S2. The recovery process is observed when the compressed lattice materials are heated above the melting point of Field's metal (see movies S1-S3). We have used a heat gun to melt the liquid metal lattice materials with temperature set as 93℃ (or 200 ℉). The normalized stress-strain curves of these three lattice materials are illustrated in Fig. 3g. Herein we define normalized stress = nominal stress / volume ratio in order to compare different lattice materials in a fair manner. It is conceived from Fig. 3g that all lattice materials exhibit a plateau stress level after yielding. The Kelvin lattice has the highest plateau stress, followed by the BCC and honeycomb structures. All stress-strain curves form closed contours since the lattice structures are restored to their original shapes after melting. Thus, we compare the normalized energy absorption and recovery ratio of these lattice materials in Fig. 3h, where the former is defined as the area enclosed by the stress-strain contour in Fig. 3g. From the quantitative comparison in Fig. 3h, we conclude that the Kelvin lattice material has the best energy absorption capacity, while the honeycomb structure is the lowest one. All three lattice materials show almost 100% recovery after re-melting, which verifies the proposed conceptual design. Note that the silicone coating has an elastic limit about 10% of strain (see Section S2). The compression testing of liquid metal lattice materials does not induce any inelastic strain in the silicone coatings in our experiment.

As shown in Fig. 4, we also tested the recovery of another two liquid metal structures: one is a soccer structure and the other is a logo "BUME" of the authors' department. Both structures are fabricated in a similar way as the lattice materials in Fig. 3. These two structures are either compressed or bended severely at room temperature. Their recoverability is assessed by re-melting the Field's metal core structures upon heating (see movies S4 and S5). The soccer is heated by a heat gun while the logo is by hot water. It is observed from the experiment that both structures were almost fully recovered. The logo shows a much faster recovery speed due to the direct heating by hot water instead of air. These two examples demonstrate broader applications of the recoverability of liquid metal structures beyond energy absorption purposes.

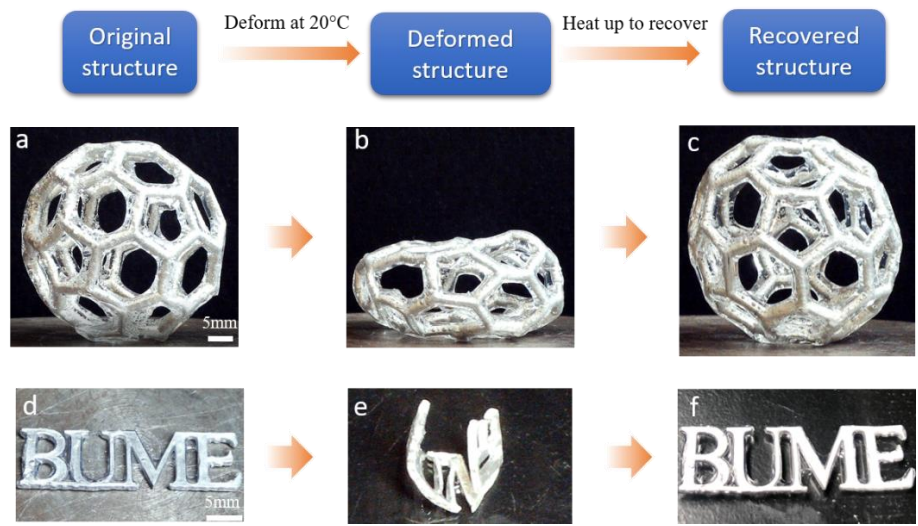

**Figure 4**. Recoverable liquid metal structures after severe deformation. (a)-(c) A liquid metal soccer. (d)-(f) A logo 'BUME'. Both of the soccer and logo structures are deformed at room temperature with severe plastic deformation and recovered above the melting point of the liquid metal.

## 4.2   Tunable shape and rigidity

Lattice materials with tunable shape and rigidity are very useful for a variety of cutting-edge applications including morphing structures in aerospace, soft robotic components, acoustic and elastic wave manipulation, among others. The phase transition of Field's metal offers a convenient way to achieve such an effect in liquid metal lattice materials. By imposing pre-strain to deform the lattice material, the shape of the liquid metal lattice materials can be fixed temporarily, and hence the rigidity of the lattice materials changes simultaneously. The pre-strain can be either imposed at room temperature by inducing plastic deformation or at high temperature followed by freezing the deformed shapes.

We have tested all the three lattice materials to evaluate how the pre-strain influences their rigidity. The experimental results are presented in Fig. 5, where Figs. 5a-c follow the same experimental procedure. Take the Kelvin lattice in Fig. 5a as an example, this lattice structure is imposed a uniaxial compression loading to 10% of strain, and held at this strain level to program the shape by melting and then solidifying the liquid metal cores. During this programming process, the stress in the liquid metal core is released, which is reflected by the abrupt decrease of the normalized stress in the stress-strain curves. After the temporary shape is fixed, we repeat the uniaxial compression for a successive 10% of strain at room temperature, and the testing is completed once the strain reaches 40% in total. All compression testing is conducted by using the MTS load frame at a loading speed of 0.1 mm/s.  Snapshots of the deformed lattice materials are illustrated in both Fig. 5 and Section S3.

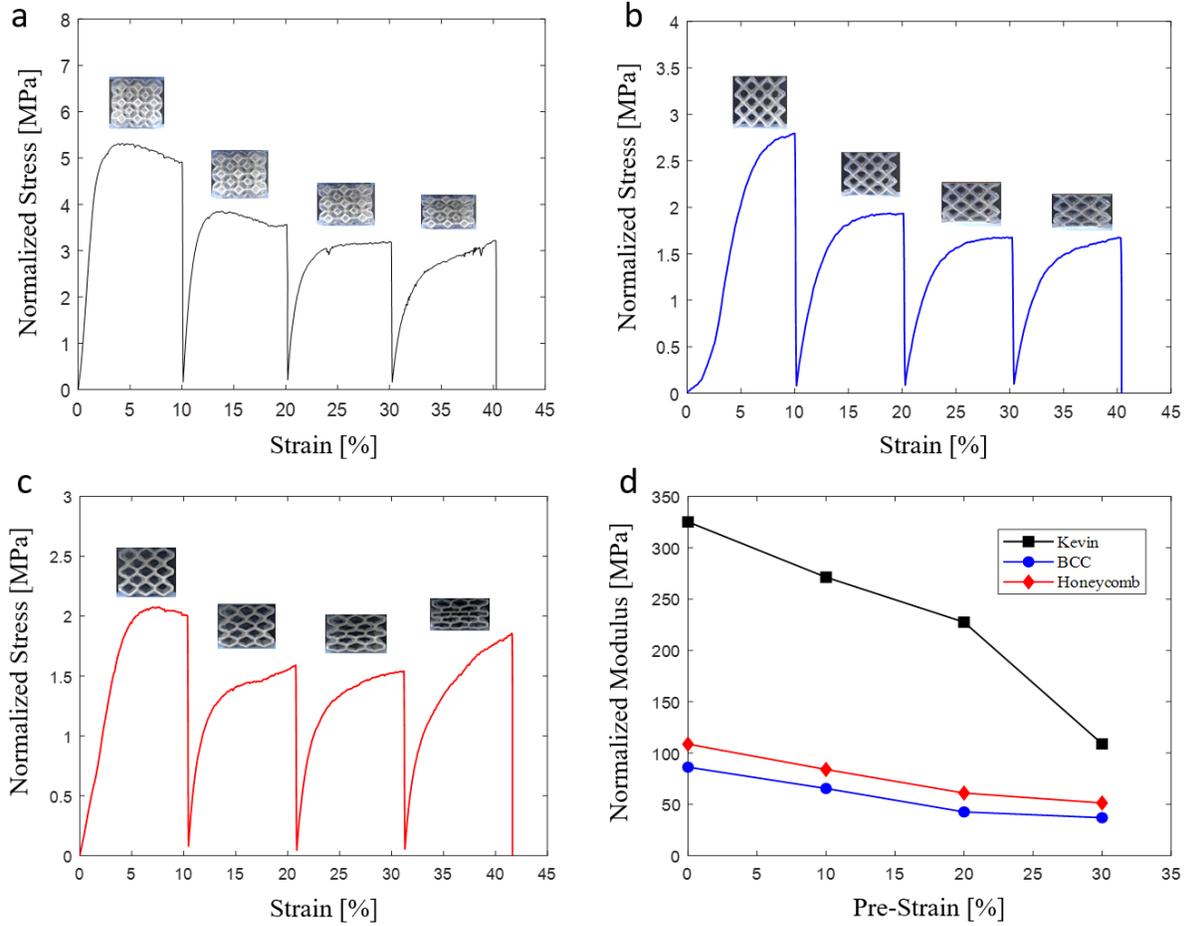

**Figure 5**. Tunable shape and rigidity of liquid metal lattice materials via imposing pre-strain. (a)-(c) Compression testing of Kelvin, BCC, and honeycomb type lattice materials with different pre-strain. (d) Rigidity of the three lattice materials with different pre-strain. The normalized modulus is calculated from the slope of the normalized stress-strain curves in the linear regimes.

Through this compression testing and re-programming process, we can evaluate the rigidity of these three lattice materials at different pre-strain levels. The normalized moduli of these lattice materials are computed from the slopes in the normalized stress-strain curves and compared in Fig. 5d. A monotonic decreasing trend is observed in Fig. 5d for the rigidity of the liquid metal lattice materials at different pre-strain levels. Among the three lattice materials, the Kelvin lattice shows the most significant rigidity declining, with the normalized modulus decreased to 1/3 of its original value once the pre-strain is 30%. This experiment demonstrates that the rigidity of liquid metal lattice materials can be tuned by imposing pre-strain. This phenomenon also has potential applications in manipulating acoustic or elastic wave propagation behaviors by using the liquid metal lattice materials, which are called tunable metamaterials [21] by some researchers.

## 4.3   Deployable and reconfigurable behaviors

Deployable structures [45] are usually designed in such a purpose that they can be stored in a very compact space and then expanded to fulfill their designed functionality. Their applications in aerospace engineering have been recognized and developed for decades; while they are also very useful in some cutting-edge technologies including soft robotics and metamaterials. The liquid metal lattice materials exhibit the deployable behavior stemming from their extrinsic shape memory effect. Once the liquid metal is melted, the elastomer shell skeleton can be deployed or deformed considerably to a compact structure and then cooled down to fix the deployed shapes. The deployed structures are then heated and restored to their original shapes during usage. Liquid metals enable the design of deployable structures with complicated geometry compared to some popular approaches in the literature [3,45,46], e.g. origami, kirigami, and mechanical mechanisms. The metallic feature of liquid metals also enable some functionalities that do not exist in shape memory polymers.

Figure 6 shows two deployable structures we fabricated, i.e. a BCC lattice structure (Fig. 6a-c) and a mesh antenna (Fig. 6d-f). Unlike the liquid metal structures in Figs. 3-5, the deployable structures are programmed at high temperature, not room temperature, which allows large volume deformation of the structures. The two deployed structures are shown in Figs. 6b and 6e and their recovered shapes are found in Figs. 6c and 6f. It is observed that the volume of the structures can be greatly reduced without affecting their recoverability.  The recovery processes can be observed in movies S6 and S7 for the BCC lattice and mesh antenna, respectively.

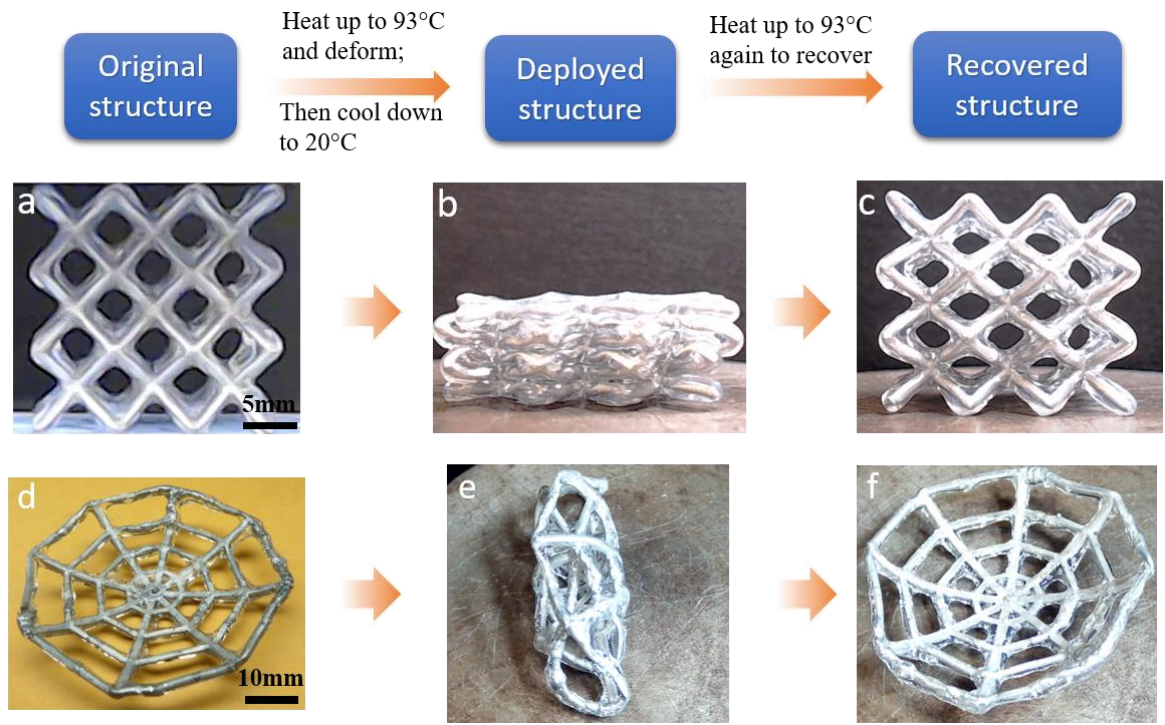

**Figure 6**. Deployable structures based on liquid metals. (a)-(c) The original, deployed, and recovered shapes of a BCC lattice structure. (d)-(e) The original, deployed, and recovered shapes of a mesh antenna.

Following the same idea, the liquid metal lattice structures can also achieve reconfigurable behaviors. An example is illustrated in Fig. 7 for a lattice hand. The fabrication process of the lattice hand follows the same procedure as described in Section 3. In Fig. 7, the lattice hand is reconfigured to two temporarily fixed shapes, namely, a love gesture (Fig. 7b) and a fist (Fig. 7d). The recovery process of these two temporarily fixed shapes are recorded in the movie S8. Certainly, by doing the programming and recovering processes repeatedly, we can reconfigure the lattice hand into unlimited gestures, which mimics a human hand. We envision that the reconfigurable behaviors of liquid metal lattice structures will open up more practical applications in the near future.

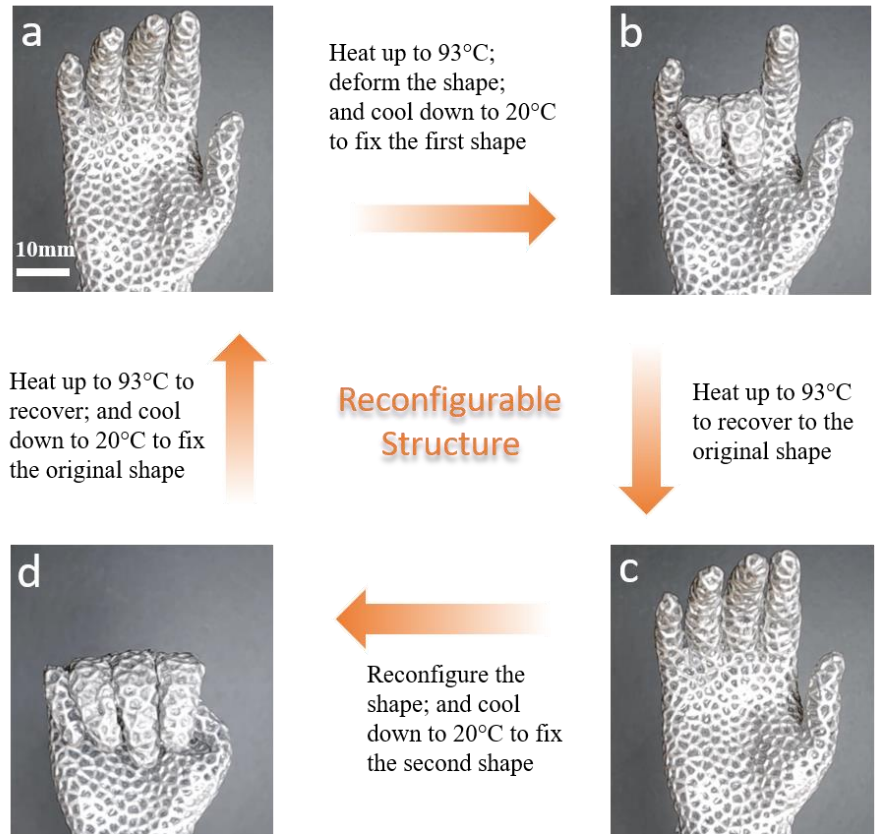

**Figure 7**. A reconfigurable lattice hand based on liquid metals. The lattice hand can be programmed and reconfigured among different temporarily fixed shapes. (a) As fabricated lattice hand. (b) Temporarily fixed shape I: Love gesture. (c) The lattice hand is restored to its original shape upon heating. (d) Temporarily fixed shape II: Fist.

## 5   Discussion and Conclusions

This work proposes a new class of shape memory lattice materials by employing liquid metals. The liquid metal lattice materials are composed of a liquid metal core and an elastomer shell skeleton. They are fabricated by a hybrid manufacturing method involving 3D printing, vacuum casting, and coating. An extrinsic shape memory effect is achieved by the thermal-induced phase transition of the liquid metals. Owning to this shape memory effect, a variety of multifunctional behaviors arise in the liquid metal lattice materials, including recoverable energy absorption, tunable shape and rigidity, deployable and reconfigurable behaviors. The liquid metal lattice materials possess synergy properties of conventional shape memory polymers and shape memory alloys. For example, they are as flexible as shape memory polymers but exhibit faster response speed, higher thermal and electrical conductivity, and narrower transition temperatures like shape memory alloys.

Future research can be towards new manufacturing processes for liquid metals, e.g. laser-based additive manufacturing, to enable fast and defect-free fabrication of liquid metal lattice materials. In addition, stronger and durable coating materials are also desirable to leverage their mechanical and multifunctional performance. Besides the three liquid metal lattice materials investigated in this work, there are a lot more lattice material types that can be manufactured and tested to assess their multifunctional properties. In addition, we envision growing applications of liquid metal lattice materials as recoverable, adaptive, and programmable components in traditional and emerging areas in the near future.

## 6    Acknowledgements

This work is supported by the start-up fund from the Waston School of Engineering and Applied Science at SUNY Binghamton. We are grateful for the discussion with Professor Junghyun Cho on the coating process.

## Appendix A

The supplementary information of this paper is accessible online.